\documentclass[acmsmall,screen,nonacm]{acmart}

\usepackage{tabularx}
\usepackage{array}

\setcopyright{none}
\renewcommand\footnotetextcopyrightpermission[1]{}
\makeatletter
\AtBeginDocument{%
  \apptocmd{\ps@firstpagestyle}{\fancyfoot{}}{}{}
  \apptocmd{\ps@standardpagestyle}{\fancyfoot{}}{}{}
  \pagestyle{standardpagestyle}%
}
\makeatother
\acmDOI{}
\acmISBN{}
\acmYear{2027}
\copyrightyear{2027}
\acmConference[FSE 2027]{ACM International Conference on the Foundations of Software Engineering}{}{}

\title[Do Code Language Models Follow Tests?]{Do Code Language Models Follow Tests? Paired Interventions on Program Behavior}
\author{Yunhao Liang}
\affiliation{%
  \institution{Chengdu Institute of Computer Applications, Chinese Academy of Sciences}
  \country{China}}
\affiliation{%
  \institution{University of Chinese Academy of Sciences}
  \country{China}}

\author{Chengguang Gan}
\affiliation{%
  \institution{Independent Researcher}
  \country{Japan}}

\author{Ruixuan Ying}
\affiliation{%
  \institution{Institute of Multidisciplinary Research for Advanced Materials (IMRAM), Tohoku University}
  \country{Japan}}

\author{Hanjun Wei}
\affiliation{%
  \institution{University of Chinese Academy of Sciences}
  \country{China}}

\author{Zhe Cui}
\affiliation{%
  \institution{Chengdu Institute of Computer Applications, Chinese Academy of Sciences}
  \country{China}}
\affiliation{%
  \institution{University of Chinese Academy of Sciences}
  \country{China}}

\author{Shiwen Ni}
\affiliation{%
  \institution{Artificial Intelligence Research Institute,\linebreak[1] Shenzhen University of Advanced Technology}
  \country{China}}
\newcommand{\QualityTasks}{180}
\newcommand{\QualityHE}{68}
\newcommand{\QualityMB}{112}
\newcommand{\QualityHighKill}{92.3}
\newcommand{\QualityLowKill}{62.0}
\newcommand{\QualityRandomKill}{91.3}

\begin{document}

\begin{abstract}
Visible tests specify concrete program behavior, but an improvement in benchmark accuracy does not establish that a model follows the rule expressed by those tests.
We study test utilization through matched prompting controls, paired semantic interventions, and test suites selected by fault detection.
Our semantic intervention holds an underspecified description and its example inputs fixed while changing the correct outputs to express one of two valid rules.
Evaluation on unseen inputs measures whether both generated programs follow their respective rules.
Across five models and three runs of 120 paired instances from 20 specification families, mean switching rates range from 11.1\% to 65.8\%. Qwen3.8-27B has the highest point estimate, followed by Qwen3.6-27B at 60.6\%; their paired difference remains uncertain.
Explicit descriptions elicit both rules from these two models on every instance, exposing a gap between implementation capability and adoption of test-specified rules.
Correct expected outputs improve MBPP+ accuracy beyond inputs alone for all five models.
On 180 tasks with fixed three-test suites, high-detection suites detect 30.3 percentage points more errors in a held-out pool dominated by reference mutants.
The corresponding generation differences range from $-0.6$ to $+1.1$ points; all intervals include zero and remain compatible with some benefit.
Paired interventions make test-specified rule changes measurable alongside implementation capability and benchmark correctness.
\end{abstract}


\maketitle

\section{Introduction}

Public tests give a code generation model concrete examples of what a program should do.
They can specify output order, boundary behavior, or tie breaking more precisely than a short natural-language description.
Test-driven prompting therefore treats examples as partial executable specifications, and prior work finds that including tests can improve generated code \citep{mathews2024tdd,fakhoury2024ticoder}.
The practical question is when a model uses the behavior those examples express.

Benchmark accuracy answers part of this question.
A correct example may resolve a missing constraint, illustrate the input format, or change which familiar solution the model retrieves.
Each route can improve pass rate.
Conversely, a model can understand an example yet fail to construct the required algorithm.
The effect of tests on final correctness consequently combines specification interpretation with implementation capability.
Requirements-clarification systems address missing information through questions and answers \citep{li2023python,mu2024clarifygpt}.
We examine the subsequent behavioral question: how reliably does a generator follow a supplied constraint when that constraint is expressed by example outputs?

We separate these outcomes with three experiments.
Matched benchmark prompts compare descriptions alone, correct input-output examples, the same inputs with incorrect outputs, and inputs alone.
The correct-versus-input-only contrast measures the contribution of expected outputs beyond concrete inputs.
Paired semantic interventions ask a more direct question: when two rules are compatible with the same description, does changing the test outputs change the generated program accordingly?
We then hold the visible-test count at three and select suites by their ability to detect errors, measuring both independent fault detection and generation accuracy.
This distinguishes a test's value for checking a program from its value for generating one.

\begin{figure}[!htbp]
\centering
\includegraphics[width=\linewidth]{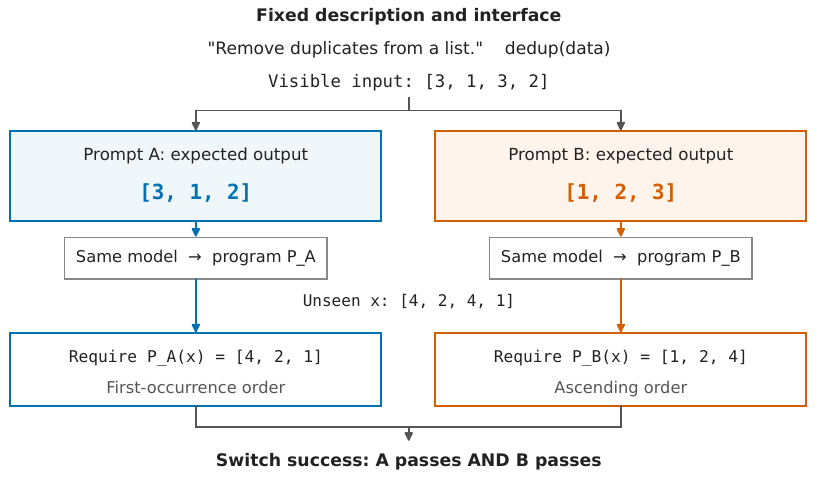}
\caption{Paired semantic intervention, illustrated with duplicate removal. The description, function interface, visible inputs, and model are fixed; only expected outputs differ between A and B. One visible and one unseen input are shown for illustration. A successful switch requires both separately generated programs to follow their assigned rules over the full held-out evaluation suite.}
\label{fig:paired-intervention}
\Description{A fixed description and visible input branch into two prompts with different expected outputs. The same model generates separate programs for A and B. Both programs must follow their assigned rules on unseen inputs for the pair to count as a successful switch.}
\end{figure}

Consider a request to remove duplicates from a list.
A program that preserves first-occurrence order and a program that sorts the distinct values both satisfy that description.
We supply the same three input lists under both conditions, with outputs that specify the corresponding order.
A successful intervention produces one program for each rule, and both must generalize to fresh inputs.
Both output sets are consistent with the same natural-language description.
Explicit descriptions of the two rules provide a separate measure of implementation capability.
Figure~\ref{fig:paired-intervention} illustrates this paired evaluation.

The intervention reveals a gap between implementing a rule and adopting it from tests.
Qwen3.6-27B and Qwen3.8-27B implement both explicitly stated rules on all paired instances across three runs, yet their mean test-conditioned switching rates are 60.6\% and 65.8\%.
Qwen3.5-9B reaches 24.2\%, while Qwen2.5 and DeepSeek each reach 11.1\%.
The 5.3-point difference between the two 27B checkpoints has a paired family interval spanning zero, so their point estimates do not establish a reliable ranking.
Separating the two directions reveals whether tests reinforce a default interpretation or induce the alternative.
Paired rule switching measures this semantic response, while benchmark pass rates measure its relationship to code correctness.

Our contributions are a paired semantic evaluation with explicit capability controls, a matched benchmark protocol with task-level inference over repeated executions, and a fixed-size test-qual\-ity experiment with separate selection and validation error pools.
These experiments connect the contribution of expected outputs to a direct measure of whether generated programs adopt the specified rule.

\section{Measuring Test Utilization}

\subsection{Matched Benchmark Prompts}

For each task, we construct four prompts with the same problem description and function interface.
\textbf{NL-only} contains the description alone.
\textbf{Correct I/O} adds public inputs and their expected outputs.
\textbf{Wrong I/O} retains those inputs and changes the expected outputs while preserving their data types.
\textbf{Inputs-only} supplies the same inputs without outputs.
We remove example blocks from the description before constructing the four conditions.
Function tasks expose up to three public examples; LiveCodeBench uses each task's official public examples.
The retained count is identical across the three example conditions for each task.

Function-task outputs are evaluated against the reference implementation.
Every correct assertion must pass, and every modified output must fail that implementation.
A value without a supported type-preserving corruption, such as \texttt{None}, remains unchanged when another output in the same suite can be changed.
We record the number of changed outputs for each task.
LiveCodeBench corruptions change the official expected stdout or return value without changing the public input or calling convention.
Thus, the wrong-output condition measures response to conflicting evidence; the paired intervention below measures response to compatible alternative specifications.
An audit of historical controls found that shuffling sometimes preserved the assertion set and that irrelevant tests changed the interface.
The matched protocol holds inputs and interfaces fixed; Section~\ref{sec:historical-analyses} reports the historical transformations and their associated analyses.

\subsection{Paired Semantic Interventions}

We author 20 specification families covering ordering, duplicate handling, tied optima, interval boundaries, arithmetic conventions, string matching, grouping, and graph reachability.
Each family defines two reference implementations, A and B, and a description that leaves their distinguishing rule open.
The function signature and input containers are explicit.
For example, a substring task receives the two-element list containing the text and pattern under both rules.

Table~\ref{tab:semantic-rules} lists the alternative rules for each family.
\begin{table}[!htbp]
\centering
\caption{Alternative rules used in the semantic intervention. Each pair shares an underspecified description, function signature, and three visible inputs. The explicit conditions add the corresponding rule to the description.}
\label{tab:semantic-rules}
\small
\begin{tabularx}{\linewidth}{@{}>{\raggedright\arraybackslash}p{.18\linewidth}>{\raggedright\arraybackslash}X>{\raggedright\arraybackslash}X@{}}
\toprule
Family & Rule A & Rule B \\
\midrule
dedup order & Preserve the order of first occurrence. & Return the distinct integers in ascending order. \\
duplicate policy & Keep the first occurrence of each value. & Keep only values that occur exactly once in the input. \\
argmax tie & Choose the first such index. & Choose the last such index. \\
threshold boundary & Include values equal to the threshold. & Include only values strictly greater than the threshold. \\
interval boundary & Include both bounds. & Include the lower bound and exclude the upper bound. \\
median choice & For even-length lists choose the lower middle value. & For even-length lists choose the upper middle value. \\
integer rounding & Round the quotient down toward negative infinity. & Truncate the quotient toward zero. \\
remainder sign & Use the nonnegative remainder associated with floor division. & Use the signed remainder associated with truncation toward zero. \\
substring overlap & Count overlapping occurrences. & Count non-overlapping occurrences from left to right. \\
substring case & Match letters case-sensitively. & Match letters case-insensitively. \\
space fields & Consecutive spaces separate empty fields; preserve leading and trailing empty fields. & Treat consecutive spaces as one separator and omit empty fields. \\
normalization & Keep decimal digits as part of the normalized text. & Keep alphabetic letters only. \\
sort ties & Preserve input order among equal scores. & Sort equal scores by ascending name. \\
partial chunk & Keep the final chunk even when it is shorter than the requested size. & Return only chunks of the full requested size. \\
group count & Group consecutive equal values into runs. & Group all occurrences of each value together. \\
graph direction & Each edge can be traversed from u to v only. & Each edge can be traversed in either direction. \\
merge touching & Also merge intervals whose endpoints touch. & Merge intervals only when their interiors overlap. \\
search index & The first position is numbered zero. & The first position is numbered one. \\
topk ties & Return exactly k values. & Include every value equal to the cutoff, even when the result has more than k elements. \\
empty segments & Keep entries that are empty after trimming. & Discard entries that are empty after trimming. \\
\bottomrule
\end{tabularx}

\end{table}

Each family has six realizations with distinct input sets, giving 120 paired instances.
We sample three visible inputs on which A and B disagree.
The A-test and B-test prompts contain these same inputs and the corresponding reference outputs.
We also generate programs from NL-only, inputs-only, and descriptions that state A or B explicitly.
The explicit conditions measure whether the model can implement the rule when it is supplied directly in language.
They provide the complete distinguishing rule as an oracle clarification, making its implementation directly observable alongside the response to tests.

Evaluation uses 30 new distinguishing inputs per instance.
For families with inputs on which both references agree, we additionally sample up to ten such inputs and require the program to pass them.
Visible and evaluation inputs are disjoint across the six realizations of a family.
We classify a program as following A or B only if it passes every input for that rule.
A program may follow neither rule.

Our primary metric is \emph{paired rule switching}: the fraction of instances for which the A-test program follows A and the B-test program follows B.
This joint requirement measures whether changing only the output constraints changes the generated behavior in the requested direction.
We also report A and B adherence separately, together with the default behavior under NL-only and inputs-only prompting.
We compute this joint outcome within each instance and run, then average the three outcomes within the instance.
Confidence intervals resample specification families, retaining all six realizations together.

\subsection{Selecting Tests by Fault Detection}

The quality experiment uses \QualityTasks{} HumanEval+/MBPP+ tasks with reference implementations.
For each task, we retain at least five distinct candidate inputs from the generated test pool and execute every candidate assertion against the reference.
The original public examples form a development set for identifying faulty programs.
We collect erroneous generated programs and single-edit AST mutants of the reference, requiring successful initialization and at least one failure on the development set.

We form a selection pool and a separate validation pool, each containing three to eight erroneous programs.
The pools use different mutation-operator families and contain no identical programs after AST normalization.
Selection mutations include arithmetic, aggregation, loop bounds, ordering parameters, and omitted filters.
Validation mutations include comparisons, constant boundaries, parameter substitutions, and input slices.
Generated programs augment these pools; their development-set behavior determines eligibility.
Across the 180 tasks, the selection pools contain 1088 erroneous programs, including 162 generated programs, and the validation pools contain 1236, including four generated programs.
The remaining programs are reference mutants.
All program sources and per-test outcomes are retained with the task.

For a suite of three tests, fault detection is the fraction of pool programs that fail at least one test in the suite.
This union criterion rewards complementary tests rather than repeatedly detecting the same error.
We enumerate the three-test combinations, restricting total assertion length to within 15\% of the median combination length.
The highest- and lowest-detection suites are selected using the selection pool, and a random suite is sampled from the same feasible combinations.
A task enters the experiment when a high-versus-low contrast exists in the selection pool.
The validation pool is then used to measure the detection rates of the selected suites; it does not determine selection or task inclusion.

We generate one program from each of four prompts: NL-only, random three tests, low-detection three tests, and high-detection three tests.
Final evaluation excludes every candidate-pool input and every development input, including candidates that were never displayed.
The two primary contrasts compare the high-detection suite with the random and low-detection suites.
We measure the detection contrast on the held-out error pool and estimate the paired generation difference on the final evaluation inputs.

\section{Experimental Setup}

\subsection{Benchmarks and Models}

The source benchmark collection contains 164 HumanEval+ tasks, 378 MBPP+ tasks, and 175 LiveCodeBench v6-new tasks.
HumanEval+ and MBPP+ extend the original benchmarks with more extensive functional tests \citep{chen2021codex,austin2021mbpp,liu2023evalplus}.
LiveCodeBench v6-new is the difference between releases v6 and v5, containing 43 easy, 52 medium, and 80 hard problems from AtCoder and LeetCode \citep{jain2024livecodebench}.
The matched protocol contains 710 tasks: 157 HumanEval+, 378 MBPP+, and all 175 LiveCodeBench tasks.
Seven HumanEval tasks lack a public example that the direct-call construction can materialize; their identifiers and reasons accompany the dataset.

All three experiments use five checkpoints: Qwen2.5-Coder-7B-Instruct \citep{hui2024qwen25coder}, Qwen3.5-9B \citep{qwen35model}, Qwen3.6-27B \citep{qwen36}, Qwen3.8-27B \citep{qwen38model}, and DeepSeek-Coder-6.7B-Instruct.
Tables abbreviate the first and last names as Qwen2.5-7B and DeepSeek-6.7B.
The study comprises 57,000 generations: 42,600 in the main experiment, 10,800 in the semantic experiment, and 3600 in the quality experiment.
All controlled generations use native chat templates, BF16 weights, greedy decoding, an 8192-token output budget, and a 16384-token context limit in vLLM 0.25.1.
Thinking is disabled for Qwen3.5, Qwen3.6, and Qwen3.8; all inputs are text.
Qwen3.8 main-experiment generation is partitioned into three request ranges with two-way tensor parallelism; the ranges retain the original batch boundaries.
The recorded execution settings accompany each experiment.
We retain complete generated solutions, including helper functions and executable entry points.

\subsection{Evaluation and Repeated Runs}

We report two correctness labels for benchmark tasks.
The complete official label uses all available benchmark tests.
The primary held-out label removes the inputs exposed through the prompt; in the quality experiment it removes the entire candidate and development input pools.
Aligned input and expected-output arrays are filtered together, while the benchmark's comparison function is preserved.
All 535 function-task references pass both versions of the main evaluator.
Function suites use a 15-second budget, extended to 60 seconds for the task whose reference repeatedly sums long integer ranges.
LiveCodeBench uses its official checker with the full suite and with private inputs that do not duplicate public inputs.
Each LiveCodeBench worker has a 4\,GiB address-space budget and a 15-second per-test time limit.
The same checker handles standard-input programs and function-call tasks.

Each main and semantic condition is executed three times, with the generation seed fixed and task order varied across runs.
These repetitions measure stability under greedy inference and batch scheduling.
For each contrast, we average its paired correctness difference over the three runs within a task.
We then resample tasks 5000 times to obtain a 95\% bootstrap interval.
The sample size remains the number of tasks: three LiveCodeBench runs yield 175 task units.
Per-run effects are reported separately, with exact McNemar tests for the task-level benchmark comparisons.
Semantic intervals resample the 20 families after averaging within instances; the quality experiment uses one generation per condition and task-level intervals.

\section{Results}

\subsection{Programs Respond to Alternative Test Rules}

Table~\ref{tab:semantic} reports means over three runs of the paired intervention.
Qwen3.8 follows rule A in 95.6\% of A-test instances and rule B in 70.3\% of B-test instances.
Its paired switching rate is 65.8\%, with a family-bootstrap 95\% interval of 47.5--83.1\%.
Qwen3.6 reaches 92.5\% A adherence, 66.9\% B adherence, and 60.6\% paired switching [43.1, 76.7].
Both checkpoints achieve 100\% adherence under explicit A and B descriptions in every run.
Their test-conditioned failures therefore occur on rules they implement when stated directly.

Qwen3.5 reaches 88.1\% A adherence and 32.5\% B adherence, with paired switching of 24.2\% [10.3, 40.0].
Its explicit A and B scores are 100.0\% and 88.6\%.
In particular, the 32.5\% B-test score is well below its 88.6\% explicit-B score, exposing a substantial gap between these ways of supplying the rule.
Qwen2.5 reaches 73.9\% A adherence and 21.1\% B adherence, with switching of 11.1\% [2.8, 21.9].
Under B tests, 65.0\% of its programs still follow A.
DeepSeek also switches in 11.1\% of instances [3.1, 21.1], with A and B adherence of 65.3\% and 31.7\%.
Its explicit scores are 68.9\% and 43.3\%, so implementation capability also limits its switching performance.

\begin{table}[!htbp]
\centering
\caption{Rule adherence and paired switching (\%) averaged over three runs of 120 paired instances from 20 specification families. Tests A/B and Explicit A/B are scored against their assigned rule. NL-only columns report the default rule distribution. Switching requires both test-conditioned programs to follow their respective rules. Intervals resample the 20 families.}
\label{tab:semantic}
\small
\begin{tabular}{lrrrrrrl}
\toprule
Model & NL: A & NL: B & Tests A & Tests B & Explicit A & Explicit B & Paired switch [95\% CI] \\
\midrule
Qwen2.5-7B & 70.8 & 10.0 & 73.9 & 21.1 & 84.4 & 65.0 & 11.1 [2.8, 21.9] \\
Qwen3.5-9B & 89.2 & 5.8 & 88.1 & 32.5 & 100.0 & 88.6 & 24.2 [10.3, 40.0] \\
Qwen3.6-27B & 70.0 & 25.0 & 92.5 & 66.9 & 100.0 & 100.0 & 60.6 [43.1, 76.7] \\
Qwen3.8-27B & 81.4 & 14.7 & 95.6 & 70.3 & 100.0 & 100.0 & 65.8 [47.5, 83.1] \\
DeepSeek-6.7B & 70.0 & 5.8 & 65.3 & 31.7 & 68.9 & 43.3 & 11.1 [3.1, 21.1] \\
\bottomrule
\end{tabular}
\end{table}

Input-only controls locate the contribution of expected outputs.
For Qwen3.8, B adherence rises from 21.1\% with inputs alone to 70.3\% with B tests, a paired increase of 49.2 points [30.3, 67.5].
For Qwen3.6, the corresponding increase is 42.8 points [23.9, 61.9].
DeepSeek's B adherence rises from 16.7\% to 31.7\%, a 15.0-point increase [5.0, 27.5].
The joint switching metric imposes a stricter requirement than these directional gains: both programs must follow their assigned rules on the same instance and run.

\begin{figure}[!htbp]
\centering
\includegraphics[width=\linewidth]{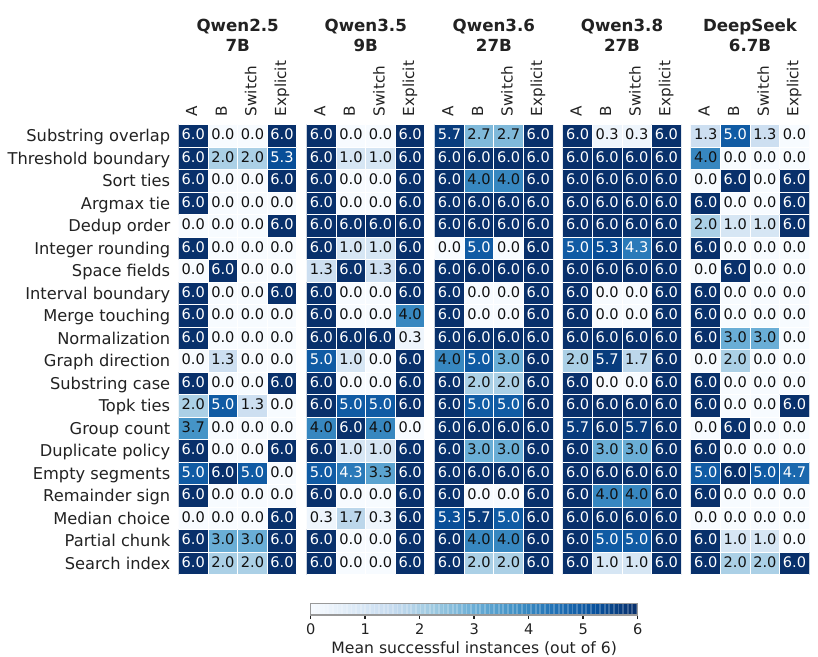}
\caption{Mean successful instances per run within each specification family, out of six. Color and cell labels show values averaged over three runs. A and B score the corresponding test-conditioned program. Switch and Explicit require successful A and B programs for the same instance and run under tests and explicit rules, respectively. All model panels use the same 0--6 scale.}
\label{fig:semantic-families}
\Description{Five annotated heatmaps show 20 specification families for Qwen2.5, Qwen3.5, Qwen3.6, Qwen3.8, and DeepSeek. Each model has A, B, Switch, and Explicit columns, with rounded means printed in every cell on a shared scale from zero to six.}
\end{figure}

The deduplication family illustrates this behavioral difference.
For one realization in the first run, Qwen2.5 generates \texttt{list(set(data))} under both test conditions, satisfying neither required order on the held-out inputs.
Qwen3.6 generates a traversal that tracks previously seen values under A tests and \texttt{sorted(set(data))} under B tests.
Figure~\ref{fig:semantic-families} gives the full family breakdown for all five models.
Qwen3.8 switches on all six realizations in every run for nine families and never switches for three; Qwen3.6 has seven such complete-success families and four complete-failure families.
Their explicit-rule controls succeed throughout these families, locating failures in the use of test-specified constraints.

Paired contrasts quantify uncertainty in the model comparisons.
Qwen3.8 has the highest mean switching rate, but its advantage over Qwen3.6 is 5.3 points with a family interval of $[-5.6,17.5]$.
Qwen3.5 exceeds Qwen2.5 by 13.1 points $[-3.3,30.8]$ and trails Qwen3.6 by 36.4 points $[21.9,51.4]$.
These comparisons do not isolate the effects of parameter count or model version, which also differ in training and architecture.

Aggregate switching rates vary little across runs (Table~\ref{tab:semantic-repeats}).
Qwen3.8 scores 65.8, 66.7, and 65.0\%; Qwen3.6 scores 60.8, 60.8, and 60.0\%; Qwen3.5 scores 24.2, 25.0, and 23.3\%.
Qwen2.5 and DeepSeek each remain between 10.8\% and 11.7\%.
The family intervals quantify variation across semantic distinctions, while these repetitions measure execution stability on the same instances.

\begin{table}[!htbp]
\centering
\caption{Per-run semantic adherence and paired switching (\%). Each run contains the same 120 instances from 20 families.}
\label{tab:semantic-repeats}
\small
\begin{tabular}{lrrrrrr}
\toprule
Model & Run & Tests A & Tests B & Explicit A & Explicit B & Paired switch \\
\midrule
Qwen2.5-7B & 1 & 74.2 & 20.0 & 84.2 & 65.0 & 10.8 \\
Qwen2.5-7B & 2 & 74.2 & 21.7 & 85.0 & 65.0 & 11.7 \\
Qwen2.5-7B & 3 & 73.3 & 21.7 & 84.2 & 65.0 & 10.8 \\
\addlinespace[2pt]
Qwen3.5-9B & 1 & 87.5 & 32.5 & 100.0 & 88.3 & 24.2 \\
Qwen3.5-9B & 2 & 89.2 & 32.5 & 100.0 & 88.3 & 25.0 \\
Qwen3.5-9B & 3 & 87.5 & 32.5 & 100.0 & 89.2 & 23.3 \\
\addlinespace[2pt]
Qwen3.6-27B & 1 & 93.3 & 66.7 & 100.0 & 100.0 & 60.8 \\
Qwen3.6-27B & 2 & 92.5 & 67.5 & 100.0 & 100.0 & 60.8 \\
Qwen3.6-27B & 3 & 91.7 & 66.7 & 100.0 & 100.0 & 60.0 \\
\addlinespace[2pt]
Qwen3.8-27B & 1 & 95.8 & 70.0 & 100.0 & 100.0 & 65.8 \\
Qwen3.8-27B & 2 & 95.8 & 70.8 & 100.0 & 100.0 & 66.7 \\
Qwen3.8-27B & 3 & 95.0 & 70.0 & 100.0 & 100.0 & 65.0 \\
\addlinespace[2pt]
DeepSeek-6.7B & 1 & 65.0 & 31.7 & 70.0 & 43.3 & 10.8 \\
DeepSeek-6.7B & 2 & 65.0 & 31.7 & 68.3 & 43.3 & 10.8 \\
DeepSeek-6.7B & 3 & 65.8 & 31.7 & 68.3 & 43.3 & 11.7 \\
\bottomrule
\end{tabular}
\end{table}

\subsection{Expected Outputs and Benchmark Correctness}

\begin{figure}[!htbp]
\centering
\includegraphics[width=\linewidth]{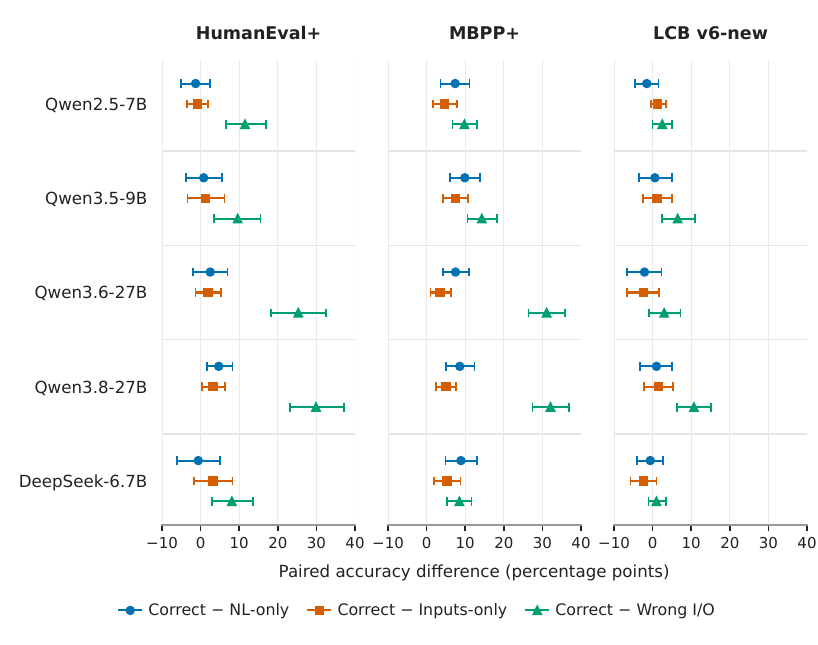}
\caption{Paired accuracy differences in percentage points with task-bootstrap 95\% intervals. Points compare correct I/O with NL-only, inputs-only, and wrong I/O on each dataset. Repeated runs are averaged within tasks before resampling. The three dataset panels compare all five models on the same horizontal scale; dashed vertical lines mark zero difference.}
\label{fig:main-effects}
\Description{Three dataset panels display 45 paired differences and their 95 percent intervals across HumanEval+, MBPP+, and LiveCodeBench. Colors and marker shapes distinguish the three prompting contrasts. Every panel uses the same scale and includes a zero reference line.}
\end{figure}

Table~\ref{tab:main} reports held-out pass rates under the matched protocol, averaged over three runs.
The contrasts in Figure~\ref{fig:main-effects} separate the contribution of expected outputs from the contribution of example inputs and from response to incorrect outputs.
Intervals use paired task differences, so tasks whose outcomes agree across conditions contribute zero to the estimated effect.
Tables~\ref{tab:official} and~\ref{tab:repeats} report the complete official scores and individual-run effects.

\begin{table}[!htbp]
\centering
\caption{Held-out pass rates (\%), averaged over three runs. Exposed inputs are removed from the evaluation suite. Each task contributes one mean correctness value.}
\label{tab:main}
\small
\begin{tabular}{llrrrrr}
\toprule
Model & Dataset & $n$ & NL-only & Correct I/O & Wrong I/O & Inputs-only \\
\midrule
Qwen2.5-7B & HumanEval+ & 157 & 82.8 & 81.5 & 70.1 & 82.4 \\
Qwen2.5-7B & MBPP+ & 378 & 66.3 & 73.7 & 63.9 & 69.0 \\
Qwen2.5-7B & LCB v6-new & 175 & 23.4 & 21.9 & 19.4 & 20.6 \\
\addlinespace[2pt]
Qwen3.5-9B & HumanEval+ & 157 & 82.8 & 83.7 & 74.1 & 82.4 \\
Qwen3.5-9B & MBPP+ & 378 & 66.2 & 76.1 & 61.8 & 68.6 \\
Qwen3.5-9B & LCB v6-new & 175 & 39.4 & 40.0 & 33.5 & 38.9 \\
\addlinespace[2pt]
Qwen3.6-27B & HumanEval+ & 157 & 91.5 & 94.1 & 68.8 & 92.1 \\
Qwen3.6-27B & MBPP+ & 378 & 72.2 & 79.7 & 48.6 & 76.2 \\
Qwen3.6-27B & LCB v6-new & 175 & 56.4 & 54.3 & 51.2 & 56.6 \\
\addlinespace[2pt]
Qwen3.8-27B & HumanEval+ & 157 & 89.6 & 94.3 & 64.3 & 91.1 \\
Qwen3.8-27B & MBPP+ & 378 & 70.9 & 79.5 & 47.4 & 74.5 \\
Qwen3.8-27B & LCB v6-new & 175 & 53.1 & 54.1 & 43.4 & 52.6 \\
\addlinespace[2pt]
DeepSeek-6.7B & HumanEval+ & 157 & 72.8 & 72.2 & 64.1 & 69.0 \\
DeepSeek-6.7B & MBPP+ & 378 & 60.4 & 69.3 & 60.8 & 64.0 \\
DeepSeek-6.7B & LCB v6-new & 175 & 14.7 & 14.1 & 13.1 & 16.4 \\
\bottomrule
\end{tabular}
\end{table}

Correct expected outputs contribute beyond input examples on MBPP+ for all five models.
The correct-minus-input-only effects are 4.7 points [1.6, 7.8] for Qwen2.5, 7.5 [4.2, 10.8] for Qwen3.5, 3.5 [1.0, 6.3] for Qwen3.6, 5.0 [2.5, 7.7] for Qwen3.8, and 5.3 [1.9, 8.8] for DeepSeek.
All five intervals exclude zero, and the effect is positive in each individual run.
For example, Qwen3.5 rises from 66.2\% with NL-only to 68.6\% with inputs alone and 76.1\% with correct I/O.
Correct expected outputs supply useful information beyond that supplied by the inputs.

HumanEval+ shows a more selective benefit.
For Qwen3.8, correct I/O reaches 94.3\%, compared with 89.6\% under NL-only and 91.1\% under inputs-only.
Its paired gains are 4.7 points [1.7, 8.3] and 3.2 points [0.4, 6.4], respectively.
The corresponding correct-minus-input-only intervals include zero for the other four models.
Qwen3.5, for example, has a 1.3-point effect $[-3.4,6.2]$, despite its recurring gain on MBPP+.
Wrong outputs lower accuracy relative to correct outputs on both function benchmarks for every model.
Qwen3.8 incurs penalties of 29.9 points on HumanEval+ and 32.1 points on MBPP+; Qwen3.6 incurs penalties of 25.3 and 31.1 points.
These two checkpoints combine high switching rates with substantial sensitivity to incorrect expected outputs.
This association does not establish that one response causes the other.

On LiveCodeBench, neither correct-minus-NL nor correct-minus-input-only intervals exclude zero for any of the five checkpoints.
The correct-minus-input-only estimates are $+1.3$, $+1.1$, $-2.3$, $+1.5$, and $-2.3$ points for Qwen2.5, Qwen3.5, Qwen3.6, Qwen3.8, and DeepSeek, respectively.
The MBPP+ benefit therefore does not extend uniformly across benchmarks, including for checkpoints with high semantic switching rates.

Individual runs expose additional variation on LiveCodeBench.
Qwen3.8's correct-minus-input-only effect is $+5.1$, $+2.3$, and $-2.9$ points, with an average of $+1.5$ points $[-2.3,5.3]$.
Qwen3.6's corresponding values are $-5.7$, $+1.1$, and $-2.3$ points.
These changing directions contrast with the positive MBPP+ effects in every run for all five models.

\begin{table}[!htbp]
\centering
\caption{Complete official-suite pass rates (\%), averaged over three runs. These include public inputs and are secondary to held-out performance.}
\label{tab:official}
\small
\begin{tabular}{llrrrrr}
\toprule
Model & Dataset & $n$ & NL-only & Correct I/O & Wrong I/O & Inputs-only \\
\midrule
Qwen2.5-7B & HumanEval+ & 157 & 82.8 & 81.5 & 69.4 & 82.4 \\
Qwen2.5-7B & MBPP+ & 378 & 66.0 & 73.5 & 63.4 & 68.8 \\
Qwen2.5-7B & LCB v6-new & 175 & 22.3 & 21.3 & 18.9 & 20.0 \\
\addlinespace[2pt]
Qwen3.5-9B & HumanEval+ & 157 & 82.2 & 83.7 & 72.6 & 81.7 \\
Qwen3.5-9B & MBPP+ & 378 & 65.8 & 75.8 & 61.6 & 68.1 \\
Qwen3.5-9B & LCB v6-new & 175 & 39.2 & 39.8 & 33.1 & 38.9 \\
\addlinespace[2pt]
Qwen3.6-27B & HumanEval+ & 157 & 91.5 & 94.1 & 66.7 & 92.1 \\
Qwen3.6-27B & MBPP+ & 378 & 72.2 & 79.7 & 48.2 & 76.2 \\
Qwen3.6-27B & LCB v6-new & 175 & 55.4 & 53.3 & 50.5 & 56.2 \\
\addlinespace[2pt]
Qwen3.8-27B & HumanEval+ & 157 & 89.6 & 94.3 & 64.3 & 91.1 \\
Qwen3.8-27B & MBPP+ & 378 & 70.8 & 79.5 & 46.2 & 74.1 \\
Qwen3.8-27B & LCB v6-new & 175 & 52.8 & 53.7 & 42.9 & 52.6 \\
\addlinespace[2pt]
DeepSeek-6.7B & HumanEval+ & 157 & 72.2 & 72.2 & 64.1 & 68.8 \\
DeepSeek-6.7B & MBPP+ & 378 & 60.1 & 68.7 & 60.1 & 63.4 \\
DeepSeek-6.7B & LCB v6-new & 175 & 14.3 & 13.7 & 12.6 & 16.0 \\
\bottomrule
\end{tabular}
\end{table}

\begin{table}[!htbp]
\centering
\caption{Per-run paired held-out differences (percentage points). Each cell lists Run 1 / Run 2 / Run 3. The same tasks are evaluated in every run; repetitions do not increase the independent sample size. Task-level gains, losses, and exact McNemar tests accompany the results.}
\label{tab:repeats}
\small
\begin{tabular}{lllll}
\toprule
Model & Dataset & Correct $-$ NL & Correct $-$ Inputs & Correct $-$ Wrong \\
\midrule
Qwen2.5-7B & HumanEval+ & -1.3 / -1.9 / -0.6 & -0.6 / -1.3 / -0.6 & +11.5 / +10.8 / +12.1 \\
Qwen2.5-7B & MBPP+ & +7.4 / +7.4 / +7.4 & +4.8 / +4.5 / +4.8 & +9.8 / +9.5 / +10.1 \\
Qwen2.5-7B & LCB v6-new & -1.1 / -0.6 / -2.9 & +1.7 / +1.7 / +0.6 & +2.9 / +2.9 / +1.7 \\
\addlinespace[2pt]
Qwen3.5-9B & HumanEval+ & +1.3 / +0.0 / +1.3 & +3.2 / -1.9 / +2.5 & +8.9 / +8.3 / +11.5 \\
Qwen3.5-9B & MBPP+ & +9.3 / +9.8 / +10.6 & +6.9 / +7.4 / +8.2 & +13.8 / +14.3 / +14.8 \\
Qwen3.5-9B & LCB v6-new & +1.7 / +1.7 / -1.7 & -1.1 / +4.0 / +0.6 & +5.7 / +4.6 / +9.1 \\
\addlinespace[2pt]
Qwen3.6-27B & HumanEval+ & +2.5 / +1.9 / +3.2 & +1.9 / +2.5 / +1.3 & +25.5 / +24.8 / +25.5 \\
Qwen3.6-27B & MBPP+ & +7.4 / +7.4 / +7.7 & +3.7 / +3.4 / +3.4 & +31.5 / +31.0 / +31.0 \\
Qwen3.6-27B & LCB v6-new & -5.7 / +0.0 / -0.6 & -5.7 / +1.1 / -2.3 & +1.7 / +4.0 / +3.4 \\
\addlinespace[2pt]
Qwen3.8-27B & HumanEval+ & +5.1 / +4.5 / +4.5 & +4.5 / +1.9 / +3.2 & +29.3 / +31.2 / +29.3 \\
Qwen3.8-27B & MBPP+ & +9.0 / +9.0 / +7.9 & +5.6 / +5.6 / +4.0 & +32.8 / +32.0 / +31.5 \\
Qwen3.8-27B & LCB v6-new & +1.7 / +4.6 / -3.4 & +5.1 / +2.3 / -2.9 & +9.1 / +12.0 / +10.9 \\
\addlinespace[2pt]
DeepSeek-6.7B & HumanEval+ & -1.3 / +0.0 / -0.6 & +3.2 / +3.8 / +2.5 & +8.3 / +9.6 / +6.4 \\
DeepSeek-6.7B & MBPP+ & +9.0 / +9.0 / +8.7 & +5.0 / +5.6 / +5.3 & +8.7 / +8.5 / +8.2 \\
DeepSeek-6.7B & LCB v6-new & -0.6 / -1.1 / +0.0 & -2.3 / -2.3 / -2.3 & +1.1 / +0.6 / +1.1 \\
\bottomrule
\end{tabular}
\end{table}

\subsection{Fault Detection and Generation Utility}

The quality dataset contains \QualityTasks{} tasks, including \QualityHE{} HumanEval+ and \QualityMB{} MBPP+ tasks.
On the held-out validation pool, high-, low-, and random-suite detection rates are \QualityHighKill\%, \QualityLowKill\%, and \QualityRandomKill\%, respectively.
These measurements quantify detection of the held-out errors, which comprise 1232 reference mutants and four model-generated programs.
Random suites can already cover most programs in the validation pool, leaving a smaller contrast against high-detection suites than against low-detection suites.
The paired high-minus-low detection difference is 30.3 percentage points, with a task-bootstrap 95\% interval of 25.3--35.3 points.
It is positive on 115 tasks, tied on 50, and negative on 15.
The corresponding high-minus-random difference is only 1.0 point, with an interval of $[-1.5,3.5]$.
Thus, the low-detection comparison supplies the substantial independent quality contrast in this experiment.

Table~\ref{tab:quality} reports generation accuracy with the same three-test budget and matched assertion lengths.
Mean total assertion lengths are 124 characters for high-detection suites, 118 for low-detection suites, and 124 for random suites.
The high-versus-low and high-versus-random comparisons measure whether fault detection translates into a useful specification for the generator.
Every outcome is evaluated on inputs outside the entire candidate pool, so passing the displayed assertions alone cannot determine the reported accuracy.

\begin{table}[!htbp]
\centering
\caption{Held-out pass rates (\%) on 180 tasks, with one run per condition and three tests per suite. Differences are percentage points with paired task-bootstrap 95\% intervals. The entire candidate and development input pools are excluded from final evaluation.}
\label{tab:quality}
\small
\begin{tabular}{lrrrrll}
\toprule
Model & NL-only & Random & Low & High & High $-$ Low & High $-$ Random \\
\midrule
Qwen2.5-7B & 69.4 & 77.2 & 74.4 & 75.6 & +1.1 [-2.2, +4.4] & -1.7 [-4.4, +1.1] \\
Qwen3.5-9B & 69.4 & 73.9 & 73.9 & 73.3 & -0.6 [-5.0, +3.9] & -0.6 [-3.9, +2.8] \\
Qwen3.6-27B & 77.8 & 83.3 & 82.2 & 82.8 & +0.6 [-3.3, +5.0] & -0.6 [-3.9, +2.8] \\
Qwen3.8-27B & 76.1 & 82.8 & 83.3 & 82.8 & -0.6 [-2.8, +1.7] & +0.0 [-3.3, +3.3] \\
DeepSeek-6.7B & 61.1 & 66.7 & 66.1 & 66.1 & +0.0 [-3.9, +3.9] & -0.6 [-3.9, +2.8] \\
\bottomrule
\end{tabular}
\end{table}

A 30.3-point detection advantage corresponds to small and uncertain generation differences across all five models.
The high-minus-low point estimates are $+1.1$, $-0.6$, $+0.6$, $-0.6$, and $0.0$ points for Qwen2.5, Qwen3.5, Qwen3.6, Qwen3.8, and DeepSeek, respectively.
For DeepSeek, both suites solve 119 of 180 tasks, with seven paired gains and seven losses.
Every interval includes zero, while the upper endpoints still permit benefits of 1.7--5.0 points.
The experiment does not establish equivalence between high- and low-detection selection.

Adding tests can nevertheless improve generation on these tasks.
High-detection suites exceed NL-only by 6.1 points [1.1, 11.1] for Qwen2.5, 5.0 [0.6, 9.4] for Qwen3.6, and 6.7 [2.2, 11.1] for Qwen3.8.
Qwen3.5 and DeepSeek have positive point estimates of 3.9 and 5.0 points, but their intervals include zero.
The gains from adding examples are distinct from the additional benefit of selecting examples for higher fault detection.
The high-minus-low results do not establish that additional benefit in the evaluated error pool.
High-minus-random intervals also include zero for all five models; random suites already detect 91.3\% of validation-pool programs.

\section{Historical Controls and Exploratory Analyses}
\label{sec:historical-analyses}

\subsection{Exploratory Synthetic LiveCodeBench Tests}
\label{sec:synthetic-lcb}

Historical synthetic LiveCodeBench tests were generated by a larger API model and checked for format and duplication.
Their expected outputs were not validated by reference implementations.
The high5 label denotes a requested budget: 118 tasks contain five tests, six contain four, 27 contain three, 18 contain two, and six contain one.
The 57 tasks below five use the pipeline's fallback handling.
The historical selector ranked surface features rather than an independently measured error-detection rate.
Table~\ref{tab:lcb-synth} reports generation outcomes for these synthetic suites.

For Qwen2.5, original relevant tests solve 26 of 175 tasks and synthetic high3 and high5 each solve 25.
For Qwen3.6, the original-test run changes from 69 to 74 solved tasks and the synthetic high5 run from 73 to 76.
The corresponding paired gains are five tasks ($p=.458$) and three tasks ($p=.701$).
The two Qwen3.6 NL-only runs use identical prompts but have only 44 identical completion strings, motivating the repeated-run protocol in the main experiment.

\begin{table}[!htbp]
\centering
\caption{Historical LiveCodeBench solved counts out of 175. Rows have separate generation runs and matched NL-only baselines. Synthetic-output correctness and fault-detection strength were not established.}
\label{tab:lcb-synth}
\begin{tabular}{llrrrr}
\toprule
Model & Test source & NL-only & Relevant & Shuffled & Irrelevant \\
\midrule
Qwen2.5-Coder-7B & Original & 23 & 26 & 24 & 27 \\
Qwen2.5-Coder-7B & Synthetic high3 & 23 & 25 & 23 & 26 \\
Qwen2.5-Coder-7B & Synthetic high5 & 23 & 25 & 24 & 27 \\
Qwen3.6-27B & Original & 69 & 74 & 78 & 68 \\
Qwen3.6-27B & Synthetic high5 & 73 & 76 & 77 & 71 \\
\bottomrule
\end{tabular}

\end{table}

\subsection{Representation and Historical Controls}

The earlier representation experiments collect prompt-end hidden states under relevant, shuffled, irrelevant, and assertion-only prompts.
These recordings describe how test context changes representations.
A control audit also establishes which behavioral comparisons those recordings support.
The historical shuffled transformation retains the complete test multiset for 160 of 378 MBPP+ tasks, 26 of 164 HumanEval+ tasks, and 9 of 175 LiveCodeBench tasks.
Irrelevant prompts copy another task's tests, including its interface.
We therefore analyze these conditions as the recorded prompt transformations; the matched and paired experiments supply the semantic comparisons.

Linear probes use five-fold cross validation grouped by task.
The best condition probes reach 82.5\% accuracy on MBPP+ and 87.7\% on HumanEval+ with Qwen2.5.
On Live\-Code\-Bench, \linebreak{}Qwen3.6's condition probe reaches 86.7\% accuracy and 0.964 macro-AUC, whereas its success probe reaches 67.7\% accuracy and 0.727 AUC.
Its rescue probe has 17 positive examples, and its best accuracy equals the 84.0\% majority baseline.
The condition label is easier to predict here.

Representation shifts measure cosine distance between each condition's prompt-end state and its NL-only state, averaged over tasks and layers.
In the historical MBPP+ experiment, adding one, two, and three synthetic tests raises mean distance from 0.0234 to 0.0262 and 0.0289.
The corresponding pass rates are 65.9\%, 68.3\%, and 68.3\%.
Assertion-only prompts shift representations further, reaching 0.0428 for the three-test setting, while removing the natural-language specification reduces correctness.
The magnitude of a representation change consequently does not determine its functional benefit.
Table~\ref{tab:shift} reports additional shift values, and Section~\ref{sec:synthetic-lcb} describes the synthetic LiveCodeBench exploration.

\subsection{Recorded Hidden-State Shifts}

Historical Qwen3.6 state collection replays the generation prompts in Transformers 5.14.1 and records the final prompt token at all 64 transformer layers.
Prompt token sequences match the vLLM generation inputs for all 1400 states in the original and synthetic-high5 runs.
Table~\ref{tab:shift} reports mean cosine distance to NL-only.
Absolute distances are interpreted within each checkpoint, since the models differ in architecture and depth.

\begin{table}[!htbp]
\centering
\caption{Historical hidden-state shifts relative to NL-only. Values describe the recorded prompts and are separate from the controlled generation experiments.}
\label{tab:shift}
\begin{tabular}{lrr}
\toprule
Condition & Peak layer & Avg. cosine distance \\
\midrule
MBPP+ original NL+tests & 17 & 0.0301 \\
MBPP+ synthetic high1 & 18 & 0.0234 \\
MBPP+ synthetic high2 & 18 & 0.0262 \\
MBPP+ synthetic high3 & 17 & 0.0289 \\
MBPP+ high3 assertion-only & 24 & 0.0428 \\
LCB synthetic high3 & 17 & 0.0105 \\
LCB synthetic high5 & 17 & 0.0120 \\
LCB high5 irrelevant & 17 & 0.0130 \\
Qwen3.6 LCB original NL+tests & 64 & 0.0185 \\
Qwen3.6 LCB original irrelevant & 64 & 0.0386 \\
Qwen3.6 LCB synthetic high5 & 64 & 0.0236 \\
Qwen3.6 LCB synthetic irrelevant & 64 & 0.0433 \\
\bottomrule
\end{tabular}

\end{table}

\section{Discussion}

The semantic intervention separates rule selection from implementing the selected rule.
A benchmark gain can reflect either component, while a paired switch requires two different programs to generalize under two corresponding sets of output constraints.
Qwen3.6 and Qwen3.8 make this distinction concrete: both implement every explicitly stated rule, yet supplying those rules through tests yields lower and asymmetric adherence.
Qwen3.5 likewise follows B in 88.6\% of explicit-B cases and only 32.5\% of B-test cases.
Qwen2.5 and DeepSeek achieve lower adherence under explicit rules, so implementation capability also contributes to their lower switching rates.

Test quality introduces a second distinction.
A test can detect a fault in a completed program without giving the generator enough information to avoid that fault.
The union of detected programs is a measurable criterion for choosing examples, and the separate validation pool measures transfer to held-out errors from different mutation families.
The generation contrasts then measure the additional step from checking power to specification utility.
These quantities should be reported together when attributing gains to stronger tests.
Here, the validation pool contains 1232 reference mutants and only four model-generated errors; transfer to a broader distribution of model failures remains an open question.
The quality experiment has one execution per condition, so it does not measure run-to-run stability as the other two experiments do.

The family-level analysis describes which semantic distinctions persist across different example inputs.
Repeated executions measure stability within those same 20 authored families, while the five checkpoints expose differences in how generators use the supplied constraints.

\section{Related Work}

\paragraph{Requirement ambiguity and clarification.}
CodeClarQA studies when to ask clarification questions and how their answers improve code generation from underspecified descriptions \citep{li2023python}.
ClarifyGPT detects ambiguity through consistency checks on generated code, asks targeted questions, and uses the answers to refine the requirement \citep{mu2024clarifygpt}.
Both approaches study acquiring missing intent and incorporating it into generation.
Our explicit-rule conditions supply that distinguishing information directly, providing an implementation control for the alternative test specifications.

Orchid evaluates how lexical, syntactic, and semantic ambiguity and vagueness affect function-level code generation \citep{yang2026assessing}.
ClarifyCodeBench evaluates interactive ambiguity resolution through clarification questions, answers, and final code, separating interaction quality from functional correctness \citep{fang2026clarifycodebench}.
These studies establish the importance of requirement specificity and clarification ability.
Our intervention fixes the underspecified description and visible inputs, changes the expected outputs between two compatible rules, and requires both resulting programs to generalize on unseen inputs.
The measured response is adoption of a supplied test constraint; the explicit prompts measure implementation of the same rule when stated in language.

\paragraph{Tests as generation context.}
Test-driven prompting studies the effect of including tests in generation prompts \citep{mathews2024tdd}.
TiCoder uses tests to formalize user intent and guide interactive generation \citep{fakhoury2024ticoder}.
The matched benchmark contrasts quantify the contribution of expected outputs beyond input examples, and the paired intervention measures directional rule changes under a fixed description.
The quality experiment additionally measures detection on held-out error programs and correctness on held-out inputs for suites with equal test counts.

\paragraph{Functional evaluation.}
HumanEval and MBPP evaluate code generation from natural-language tasks \citep{chen2021codex,austin2021mbpp}.
EvalPlus expands their evaluation tests and exposes errors missed by the original suites \citep{liu2023evalplus}.
LiveCodeBench evaluates models on programming problems collected over time \citep{jain2024livecodebench}.
Our benchmark contrasts preserve these evaluators while separating exposed inputs from final evaluation inputs.

\section{Conclusion}

Tests can change the semantic rule followed by a generated program, and the strength of this response varies sharply between the evaluated checkpoints.
Across three runs, mean paired switching rates range from 11.1\% to 65.8\% over five models.
Qwen3.8 has the highest point estimate, but its paired contrast with Qwen3.6 does not establish a clear advantage.
Both models implement every explicitly stated rule while following test-specified rules less reliably.

Matched benchmark controls, independent test-quality measurements, and task-level repeated-run analysis connect rule following to generation correctness.
Correct outputs contribute beyond input examples on MBPP+ for all five models and on HumanEval+ for Qwen3.8.
In the quality experiment, a 30.3-point detection advantage on a held-out pool dominated by reference mutants corresponds to uncertain generation differences ranging from $-0.6$ to $+1.1$ points.
Paired rule switching makes the semantic effect of expected outputs observable alongside these correctness measurements.

\section*{Data Availability}
The code and experimental data will be provided to reviewers through an anonymous repository at \url{https://anonymous.4open.science/r/do-code-models-follow-tests-C7EA}.

\bibliographystyle{ACM-Reference-Format}
\bibliography{paper}

\end{document}